\documentclass[eqsecnum,superscriptaddress,nofootinbib,11pt]{revtex4}
\usepackage{amsmath}
   \usepackage{bm}
\usepackage{amsthm,amscd}
\usepackage{mathrsfs}
\usepackage{verbatim}
\usepackage{amsfonts,amsmath,latexsym,amssymb,txfonts,bm}
\usepackage[american]{babel}
\usepackage{dsfont}
\usepackage[dvips]{graphicx}
\usepackage{latexsym}
\usepackage{epsfig}
\newcommand{\diff}{\mathrm{d}}

\def\beq{\begin{eqnarray}}
\def\eeq{\end{eqnarray}}

\begin{document}
\title{Some new results for the one-loop mass correction to the compactified $\lambda\phi^{4}$ theory}

\author{Guglielmo Fucci\footnote{Electronic address: fuccig@ecu.edu}}
\affiliation{Department of Mathematics, East Carolina University, Greenville, NC 27858 USA}

\author{Klaus Kirsten\footnote{Electronic address: Klaus\textunderscore Kirsten@Baylor.edu}}
\affiliation{GCAP-CASPER, Department of Mathematics, Baylor University, Waco, TX 76798 USA}

\date{\today}
\vspace{2cm}
\begin{abstract}

In this work we consider the one-loop effective action of a self-interacting $\lambda\phi^{4}$ field propagating in a $D$ dimensional Euclidean space
endowed with $d\leq D$ compact dimensions. The main purpose of this paper is to compute the corrections to the mass of the field due to the presence of the
compactified dimensions. Although results for the one-loop correction to the mass of a $\lambda\phi^{4}$ field are very well known for compactified toroidal spaces,
where the field obeys periodic boundary conditions, similar results do not appear to be readily available for cases in which the scalar field is subject
to Dirichlet and Neumann boundary conditions. We apply the results for the one-loop mass correction to the study of the critical temperature in Ginzburg-Landau models.

\end{abstract}
\maketitle

\section{Introduction}

One of the simplest, and most heavily studied, models describing a self-interacting quantum field is provided by the $\lambda\phi^{4}$ theory \cite{ryder}.
This theory allows for the analysis, among other things, of important processes such as symmetry breaking and symmetry restoration. Although $\lambda\phi^{4}$ theories
have been originally developed for fields propagating in the (infinite) Minkowski space, a significant amount of work has been performed in the past thirty years which
focuses on the study of $\lambda\phi^{4}$ theories on compactified spaces. The reasons for considering field propagation in compact spaces are manifold.
For instance, quantum field theory at finite temperature can be obtained, via the Matsubara formalism, by compactifying one dimension to a circle of radius equal
to the inverse of the temperature \cite{kapusta}.

A second reason for considering compactified spaces can be found in the fact that
the structure of the vacuum in quantum field theory is sensitive to the topology of the space in which the field propagates.
This implies that the Casimir energy, which is the energy associated with the vacuum, depends on the way the space is compactified and on the type of boundary
conditions that the field must obey \cite{amb83,blau88,bytse03,cognola92, elizalde94a,elizalde95}.

When a quantum field is constrained in a compact space not only its vacuum energy but also all other characteristics
of the field including, in particular, its mass are modified \cite{toms80}. In fact, the mass term in the effective potential acquires, in compactified spaces, 
a contribution which is dependent on the topology of the compactified dimensions. Such a term, which is known as the topological mass, could, for example, 
break a specific symmetry of the theory giving rise to the phenomenon of topological symmetry breaking. Compactified spaces have lately attracted the attention of researchers
interested in studying the effect of compactified dimensions on the critical temperature of fields undergoing a phase transition in Ginzburg-Landau models \cite{khanna14}.
These studies are very important, in particular, in order to gain a deeper understanding of phase transitions in superconductors and how the temperature at which the transition occurs depends on the geometry of the superconductor \cite{abreu05}.

While the one-loop mass corrections to the self-interacting $\lambda\phi^{4}$ theory have been computed by several authors throughout the years in the case of periodic boundary conditions \cite{actor88,actor90,elizalde94}, to our knowledge, the cases of Dirichlet and Neumann boundary conditions have not received the same attention in the literature. This is, perhaps, due to the more complicated formalism involved in the Dirichlet and Neumann cases compared to the one encountered in the periodic case. Moreover, considering compactified spaces with periodic boundary conditions means, in fact, limiting the analysis to those spaces possessing a toroidally compactified subspace. By allowing the scalar field to obey Dirichlet
and Neumann boundary conditions we actually extend the results for the mass corrections to other types of compactified spaces which have not been previously considered.

One of the goals of this paper is to provide explicit results for the mass corrections when the scalar field propagates in a
compactified space endowed with Dirichlet and Neumann boundary conditions.
We employ the spectral zeta function formalism in order to obtain an expression for the regularized one loop effective potential of the $\lambda\phi^{4}$ theory
and, hence, for the one-loop corrections to the mass of the field. The expressions for the mass
corrections obtained by considering Dirichlet, Neumann, and periodic boundary conditions are then renormalized by using the heat kernel asymptotic expansion method.
In addition to providing explicit results for the one-loop mass corrections, we also exploit them to obtain, in the ambit of Ginzburg-Landau models, equations
describing how the critical temperature at which a phase transition occurs, depends on the geometric properties of the compactified subspace. Our results on the critical
temperature complement and expand those obtained in \cite{abreu05,malbo02}

The outline of the paper is the following. In the next section we write an expression for the one-loop effective action for the $\lambda\phi^{4}$ theory
in terms of the spectral zeta function. From the one-loop effective action, we then derive explicit expressions for the one-loop corrections to the mass of the field
when periodic, Dirichlet and Neumann boundary conditions are imposed on the $d$-dimensional subspace. The results for the mass corrections
are then used to analyze how the critical temperature in the Ginzburg-Landau model depends on the size of the $d$-dimensional subspace. The conclusions
summarize the main results of the paper and point to a few directions for further study.

\section{One-loop effective potential and the spectral zeta function}

For our analysis we consider the following $D$-dimensional Euclidean space
\begin{equation}
{\cal M}=\mathbb{R}^{D-d}\times [0,L_{1}]\times \cdots\times[0,L_{d}]\;,
\end{equation}
where the topological product of the line intervals $[0,L_{i}]$ represents the $d$-dimensional compactified subspace of ${\cal M}$. To describe the
$\lambda\phi^{4}$ theory we employ the well-known Hamiltonian functional
\begin{equation}\label{1}
H=\frac{1}{2}(\partial_{\mu}\phi)(\partial^{\mu}\phi)+\frac{1}{2}m^{2}\phi^{2}+\frac{\lambda}{4!}\phi^{4}\;,
\end{equation}
where $\phi$ represents the scalar field of mass $m$ propagating in the space ${\cal M}$ defined above. By rewriting the scalar field
as a sum of a classical constant background field $\Phi$ and a quantum fluctuation $\varphi$, as $\phi=\Phi+\varphi$, one can show that
$\varphi$ satisfies the following differential equation
\begin{equation}\label{2}
(-\Delta+M^{2})\varphi=0\;,
\end{equation}
where the modified mass $M^{2}$ is defined according to the formula $M^{2}=m^{2}+(1/2)\lambda\Phi^{2}$. In this framework the effective potential
can be found to be
\begin{equation}\label{3}
V_{\textrm{eff}}=\frac{1}{2}m^{2}\Phi^{2}+\frac{\lambda}{4!}\Phi^{4}+U(M,\mu)\;,
\end{equation}
where $U(M,\mu)$ denotes the one-loop effective potential per unit volume
\begin{equation}\label{4}
V_{d}V_{D-d}U(M,\mu)=\frac{1}{2}\ln\textrm{Det}\left(\frac{-\Delta+M^{2}}{\mu^{2}}\right),
\end{equation}
where $V_{d}$ is the volume of the compactified $d$-dimensional subspace and
$V_{D-d}$ denotes the unit volume for the remaining $D-d$ dimensions, and $\mu$ is a parameter with
the dimension of mass. The functional determinant in (\ref{4}) can be defined in terms of the spectral zeta function of the
operator $-\Delta+M^{2}$ \cite{elizalde94a,kirsten01} and, hence, the one-loop effective potential takes the form
\begin{equation}\label{5}
V_{d}U(M,\mu)=-\frac{1}{2}[\zeta(0,M)\ln\mu^{2}+\zeta'(0,M)]\;,
\end{equation}
where we have performed the limit $V_{D-d}\to\infty$ to account for the unconstrained $D-d$ dimensions of ${\cal M}$. The spectral zeta function {\it density} in (\ref{5})
is defined as
\begin{equation}\label{6}
\zeta(s,M)=\frac{1}{(2\pi)^{D-d}}\sum_{n}\int_{\mathbb{R}^{D-d}}\left[\nu_{n}^{2}+|k|^{2}+M^{2}\right]^{-s}\diff^{D-d}k\;,
\end{equation}
where $\nu_{n}$ are the eigenvalues of the Laplace operator $-\Delta$ on the compactified dimensions. 
By performing the integration over the variable $k$ one obtains
\begin{equation}\label{6a}
\zeta(s,M)=\frac{1}{(4\pi)^{\frac{D-d}{2}}}\frac{\Gamma\left(s-\frac{D-d}{2}\right)}{\Gamma(s)}\sum_{n}\left(\nu_{n}^{2}+M^{2}\right)^{-s+\frac{D-d}{2}}\;.
\end{equation}
The specific form of the eigenvalues $\nu_{n}$ depends on the particular boundary condition
one imposes on the field in the compactified subspace of ${\cal M}$. As we have mentioned earlier, the vast majority of the authors in the literature
consider toroidally compactified subspaces which lead to periodic boundary conditions imposed on the field. Here, we consider not only periodic,
but also Dirichlet and Neumann  boundary conditions.
By imposing Dirichlet boundary conditions on the field in the $d$-dimensional subspace of ${\cal M}$ the operator $-\Delta$ has the eigenvalues
\begin{equation}\label{7}
\nu^{2}_{n}=\sum_{j=1}^{d}\frac{n_{j}^{2}\pi^{2}}{L_{j}^{2}}\;,\quad (n_{1},n_{2},\ldots, n_{d})\in\mathbb{N}^{d}_{+}\;,
\end{equation}
for Neumann boundary conditions one obtains, instead,
\begin{equation}\label{8}
\nu^{2}_{n}=\sum_{j=1}^{d}\frac{n_{j}^{2}\pi^{2}}{L_{j}^{2}}\;,\quad (n_{1},n_{2},\ldots, n_{d})\in\mathbb{N}^{d}_{0}\;,
\end{equation}
and finally when periodic boundary conditions are imposed one has
\begin{equation}\label{9}
\nu^{2}_{n}=\sum_{j=1}^{d}\frac{4n_{j}^{2}\pi^{2}}{L_{j}^{2}}\;,\quad (n_{1},n_{2},\ldots, n_{d})\in\mathbb{Z}^{d}\;.
\end{equation}
By using the explicit eigenvalues in (\ref{7}), (\ref{8}), and (\ref{9}) in the expression (\ref{6a}) we have
\begin{equation}\label{10}
\zeta_{i}(s,M)=\frac{1}{(4\pi)^{\frac{D-d}{2}}}\frac{\Gamma\left(s-\frac{D-d}{2}\right)}{\Gamma(s)}{\cal Z}_{i}\left(s-\frac{D-d}{2},M\Bigg|L_{1},\ldots, L_{d}\right)\;,
\end{equation}
where $i=\{D,N,P\}$ indicates whether we are considering Dirichlet, Neumann, or periodic boundary conditions. The newly introduced functions ${\cal Z}_{i}$
are defined in terms of the sums
\begin{equation}\label{11}
{\cal Z}_{D}(s,M|L_{1},\ldots,L_{d})=\sum_{n_{1},\ldots,n_{d}=1}^{\infty}\left[ \sum_{j=1}^{d}\frac{n_{j}^{2}\pi^{2}}{L_{j}^{2}}+M^{2}\right]^{-s}\;,\quad
{\cal Z}_{N}(s,M|L_{1},\ldots,L_{d})=\sum_{n_{1},\ldots,n_{d}=0}^{\infty}\left[ \sum_{j=1}^{d}\frac{n_{j}^{2}\pi^{2}}{L_{j}^{2}}+M^{2}\right]^{-s}\;,
\end{equation}
and for periodic boundary conditions
\begin{equation}\label{12}
{\cal Z}_{P}(s,M|L_{1},\ldots,L_{d})=\sum_{n_{1},\ldots,n_{d}=-\infty}^{\infty}\left[ \sum_{j=1}^{d}\frac{4n_{j}^{2}\pi^{2}}{L_{j}^{2}}+M^{2}\right]^{-s}\;.
\end{equation}
The infinite sums appearing in (\ref{11}) and (\ref{12}) can be expressed in terms of the Epstein zeta function \cite{elizalde95,epst03-56-615,epst07-63-205,kirsten10}
\begin{equation}\label{13}
\zeta_{E}(s,c|{\bf r})=\sum_{{\bf m}\in\mathbb{Z}^{d}}\left(c+r_{1}m_{1}^{2}+r_{2}m_{2}^{2}+\ldots +r_{d}m_{d}^{2}\right)^{-s}\;,
\end{equation}
with $\Re (s)>d/2$, $c\in\mathbb{R}^{+}$, and ${\bf r}\in\mathbb{R}_{d}^{+}$
as follows \cite{amb83}: we consider the function ${\cal Z}_{D}$ and utilize the Mellin transform to write
\begin{equation}\label{14}
{\cal Z}_{D}(s,M|L_{1},\ldots,L_{d})=\frac{1}{\Gamma(s)}\int_{0}^{\infty}t^{s-1}\prod_{j=1}^{d}\left(\sum_{n_{j}=1}^{\infty}e^{-t\frac{\pi^{2}}{L_{j}^{2}}n_{j}^{2}}\right)
e^{-tM^{2}}\diff t\;.
\end{equation}
By expressing the sum in terms of the Jacobi theta function \cite{gradshteyn}
\begin{equation}\label{15}
\theta_{3}(z,q)=\sum_{n=-\infty}^{\infty}q^{n^{2}}\cos(2nz)=1+2\sum_{n=1}^{\infty} q^{n^{2}}\cos(2nz)\;,
\end{equation}
for $|q|<1$, we obtain
\begin{equation}\label{16}
{\cal Z}_{D}(s,M|L_{1},\ldots,L_{d})=\frac{1}{2^{d}\Gamma(s)}\int_{0}^{\infty}t^{s-1}\prod_{j=1}^{d}\left[-1+\theta_{3}\left(0,e^{-t\frac{\pi^{2}}{L_{j}^{2}}}\right)\right]
e^{-tM^{2}}\diff t\;.
\end{equation}
The product appearing in the integrand of (\ref{16}) can be written as
\begin{equation}\label{17}
\prod_{j=1}^{d}\left[-1+\theta_{3}\left(0,e^{-t\frac{\pi^{2}}{L_{j}^{2}}}\right)\right]=(-1)^{d}+\sum_{l=1}^{d}(-1)^{d-l}\sum_{1\leq i_{1}<i_{2}<\cdots<i_{l}\leq d}\prod_{k=1}^{l}\theta_{3}\left(0,e^{-t\frac{\pi^{2}}{L_{i_{k}}^{2}}}\right)\;,
\end{equation}
and hence, by using also the definition (\ref{15}), the function in (\ref{16}) becomes
\begin{eqnarray}\label{18}
\lefteqn{{\cal Z}_{D}(s,M|L_{1},\ldots,L_{d})=\frac{(-1)^{d}}{2^{d}}M^{-2s}}\nonumber\\
&&+\sum_{l=1}^{d}(-1)^{d-l}\sum_{1\leq i_{1}<i_{2}<\cdots<i_{l}\leq d}\left[\sum_{n_{i_{1}},\ldots,n_{i_{l}}=-\infty}^{\infty}
\int_{0}^{\infty}t^{s-1}e^{-t\left(\frac{\pi^{2}}{L^{2}_{i_{1}}}n_{i_{1}}^{2}+\cdots+\frac{\pi^{2}}{L^{2}_{i_{l}}}n_{i_{l}}^{2}+M^{2}\right)}\diff t\right]\;.
\end{eqnarray}
Once the simple integral in (\ref{18}) is computed and the definition of the Epstein zeta function in (\ref{13}) is invoked, one finally obtains an expression for
${\cal Z}_{D}(s,M|L_{1}\ldots,L_{d})$ as a linear combination of Epstein zeta functions, namely
\begin{eqnarray}\label{19}
{\cal Z}_{D}(s,M|L_{1},\ldots,L_{d})=\frac{(-1)^{d}}{2^{d}}M^{-2s}+\frac{1}{2^{d}}\sum_{l=1}^{d}(-1)^{d-l}\sum_{1\leq i_{1}<i_{2}<\cdots<i_{l}\leq d}\zeta_{E}\left(s,M^{2}\Bigg|\frac{\pi^{2}}{L^{2}_{i_{1}}},\ldots,\frac{\pi^{2}}{L^{2}_{i_{l}}}\right)\;.
\end{eqnarray}

An argument similar to the one outlined for ${\cal Z}_{D}(s,M|L_{1},\ldots,L_{d})$ can be applied to the series appearing in the definition
of ${\cal Z}_{N}(s,M|L_{1},\ldots,L_{d})$. By using the Mellin transform and by noticing that
\begin{equation}
\sum_{n_{j}=0}^{\infty}e^{-t\frac{\pi^{2}}{L_{j}^{2}}n_{j}^{2}}=\frac{1}{2}\left[1+\theta_{3}\left(0,e^{-t\frac{\pi^{2}}{L_{j}^{2}}}\right)\right]\;,
\end{equation}
we obtain, by following the same steps performed for the Dirichlet case, the expression
\begin{equation}\label{20}
{\cal Z}_{N}(s,M|L_{1},\ldots,L_{d})=\frac{M^{-2s}}{2^{d}}+\frac{1}{2^{d}}\sum_{l=1}^{d}\sum_{1\leq i_{1}<i_{2}<\cdots<i_{l}\leq d}\zeta_{E}\left(s,M^{2}\Bigg|\frac{\pi^{2}}{L^{2}_{i_{1}}},\ldots,\frac{\pi^{2}}{L^{2}_{i_{l}}}\right)\;.
\end{equation}
The case involving periodic boundary conditions is actually the simplest one. In fact, ${\cal Z}_{P}(s,M|L_{1},\ldots,L_{d})$ in (\ref{12})
can be directly written in terms of the Epstein zeta function as follows
\begin{equation}\label{21}
{\cal Z}_{P}(s,M|L_{1},\ldots,L_{d})=\zeta_{E}\left(s,M^{2}\Bigg|\frac{4\pi^{2}}{L^{2}_{i_{1}}},\ldots,\frac{4\pi^{2}}{L^{2}_{i_{d}}}\right)\;.
\end{equation}
Let us point out that the expressions for ${\cal Z}_{D}(s,M|L_{1}\ldots,L_{d})$ and ${\cal Z}_{N}(s,M|L_{1},\ldots,L_{d})$, and hence for the corresponding
spectral zeta functions, are much more involved than the one for ${\cal Z}_{P}(s,M|L_{1},\ldots,L_{d})$. This is, perhaps, the main reason why
the one-loop mass corrections to the self-interacting $\lambda\phi^{4}$ theory in compact spaces have been considered in the literature almost exclusively
for periodic boundary conditions.

The spectral zeta function in (\ref{10}) is defined for $\Re(s)>D/2$ and it can be analytically continued to a meromorphic function in the entire complex plane 
\cite{elizalde95,epst03-56-615,epst07-63-205,kirsten01}.
The analytic continuation of $\zeta_{i}(s,M)$ is directly provided by the analytic continuation of the Epstein zeta function.

\section{The one-loop mass correction}

For each of the three boundary conditions considered in this work we have the one-loop effective potential (cf. (\ref{5}))
\begin{equation}\label{22}
V_{d}U_{i}(M,\mu)=-\frac{1}{2}\zeta_{i}(0,M)\ln\mu^{2}-\frac{1}{2}\zeta'_{i}(0,M)\;,
\end{equation}
where we can define $V_{d}=L_{1}\cdots L_{d}$.
According to the expression in (\ref{3}) the one-loop correction to the mass of the field, $\delta m_{i}^{2}$, is proportional to the
coefficient of $\Phi^{2}$ in (\ref{22}).
Our next task therefore consists in isolating the terms in (\ref{22}) containing $\Phi^{2}$.
This can be accomplished by using, once again, the Mellin transform to rewrite (\ref{6}), for $\Re(s)>D/2$, as
\begin{equation}\label{23}
\zeta_{i}(s,M)=\frac{1}{(2\pi)^{D-d}}\sum_{n}\int_{\mathbb{R}^{D-d}}\left(\frac{1}{\Gamma(s)}\int_{0}^{\infty}t^{s-1}e^{-t(\nu_{n}^{2}+|k|^{2}+M^{2})}\diff t\right)\diff^{D-d}k\;.
\end{equation}
In the integrand we substitute the explicit expression of $M$ in terms of $\Phi^{2}$, namely $M^{2}=m^{2}+(1/2)\lambda\Phi^{2}$, and we expand the
resulting formula to obtain
\begin{eqnarray}\label{24}
\zeta_{i}(s,M)&=&\frac{1}{(2\pi)^{D-d}}\sum_{n}\int_{\mathbb{R}^{D-d}}\left[\frac{1}{\Gamma(s)}\int_{0}^{\infty}t^{s-1}e^{-t(\nu_{n}^{2}+|k|^{2}+m^{2})}\diff t\right]\diff^{D-d}k\nonumber\\
&-&\frac{\lambda\Phi^{2}}{2(2\pi)^{D-d}}\sum_{n}\int_{\mathbb{R}^{D-d}}\left[\frac{1}{\Gamma(s)}\int_{0}^{\infty}t^{s}e^{-t(\nu_{n}^{2}+|k|^{2}+m^{2})}\diff t\right]\diff^{D-d}k+O\left(\Phi^{4}\right)\;.
\end{eqnarray}
By comparing the integrals in (\ref{24}) with the representation of the spectral zeta function in (\ref{23}) it is not difficult to obtain the relation
\begin{equation}\label{25}
\zeta_{i}(s,M)=\zeta_{i}(s,m)-\frac{\lambda}{2}\Phi^{2}s\zeta_{i}(s+1,m)+O\left(\Phi^{4}\right)\;.
\end{equation}
By substituting (\ref{25}) in (\ref{22}) and by collecting all the terms proportional to $\Phi^{2}$ one arrives
at the following expression for the one-loop mass correction
\begin{equation}\label{26}
V_{d}\delta m^{2}_{i}=\frac{\lambda}{2}\lim_{s\to 0}\left[s\zeta_{i}(s+1,m)\ln\mu^{2}+\zeta_{i}(s+1,m)+s\zeta'_{i}(s+1,m)\right]\;.
\end{equation}
According to the general theory of the spectral zeta function \cite{kirsten01,seel68-10-288,voros87}, $\zeta_{i}(s,m)$ is a meromorphic function
possessing simple poles located at the points $s=(D-k)/2$ with $k=\{0,\ldots, D-1\}$ and $s=-(2l+1)/2$ for $l\in\mathbb{N}_{0}$. This implies that $\zeta_{i}(s+1,m)$
will generally develop a pole at $s=0$ and can be expanded in a Laurent series as
\begin{equation}\label{27}
\zeta_{i}(s+1,m)=\frac{1}{s}\textrm{Res}\;\zeta_{i}(1,m)+\textrm{FP}\;\zeta_{i}(1,m)+O(s)\;,
\end{equation}
where $\textrm{Res}$ and $\textrm{FP}$ denote, respectively, the residue and the finite part.

The expansion obtained in (\ref{27}) can be used to evaluate the limit in (\ref{26}) to finally find the following remarkably simple expression for the one-loop mass correction
\begin{equation}\label{28}
V_{d}\delta m^{2}_{i}=\frac{\lambda}{2}\left[\textrm{Res}\;\zeta_{i}(1,m)\ln\mu^{2}+\textrm{FP}\;\zeta_{i}(1,m)\right]\;.
\end{equation}
The above formula for $\delta m^{2}_{i}$ depends explicitly on the arbitrary mass parameter $\mu$ and, hence, needs to be renormalized \cite{toms80}.
The renormalization process is relatively simple when one considers periodic boundary conditions. In fact, in this case one can show that the quantity
$V_{d}^{-1}\textrm{Res}\;\zeta_{i}(1,m)$ does not depend on the compactification lengths of the torus \cite{elizalde94} and the renormalization can be
performed by simply taking the limit as the compactification lengths go to infinity. When one considers other types of boundary conditions,
such as Dirichlet or Neuman, the topology of the space differs from a torus and, therefore, the coefficient of $\ln\mu^{2}$ in (\ref{28})
can, in general, depend on the compactification lengths $L_{i}$ and the limit will simply not give a renormalized expression for the one-loop mass correction.
To obtain a renormalized expression for $\delta m^{2}_{i}$ which is valid for the three types of boundary conditions we are studying, we have to exploit
the heat kernel asymptotic expansion \cite{birrel,bunch,cognola94,toms82,vassil}. The counter-terms needed to renormalize the
one-loop effective action and, hence, the mass correction are proportional to the first suitable number of terms of the heat
kernel asymptotic expansion.

In order to find the required counter-terms for the mass correction we consider the well-known relation
between the spectral zeta function of the operator $-\Delta+m^{2}$ and the ${\mathscr L}^{2}$-trace of the associated heat kernel 
\begin{equation}\label{29}
\zeta_{i}(s,m)=\frac{1}{\Gamma(s)}\int_{0}^{\infty}t^{s-1}\textrm{Tr}_{\mathscr{L}^{2}}\left[e^{-t(-\Delta+m^{2})}\right]\diff t\;,
\end{equation}
valid for $\Re(s)>D/2$.
The representation (\ref{29}) is then used to rewrite the terms in the expression (\ref{26}) in the form of integrals. The trace of the heat
kernel that appears in each integral is then replaced with the resummed form of its small-$t$ asymptotic expansion (see e.g. \cite{vassil})
\begin{equation}\label{30}
\textrm{Tr}_{\mathscr{L}^{2}}\left[e^{-t(-\Delta+m^{2})}\right]\sim\frac{1}{(4\pi t)^{\frac{D}{2}}}e^{-tm^{2}}\sum_{k=0}^{\infty}A^{(i)}_{\frac{k}{2}}t^{\frac{k}{2}}\;,
\end{equation}
where the heat kernel expansion coefficients $A^{(i)}_{k/2}$ are universal functions of geometric invariants \cite{mina49,gilkey95}. The resulting
elementary integrals can be computed to obtain
\begin{eqnarray}\label{31}
V_{d}\delta m^{2}_{i}&\sim& \frac{\lambda}{2(4\pi)^{\frac{D}{2}}}\sum_{j=0}^{\left[\frac{D}{2}-1\right]}\frac{(-1)^{j}}{j!}A^{(i)}_{\frac{D}{2}-j-1}m^{2j}\left[H_{j}-\ln\left(\frac{m^{2}}{\mu^{2}}\right)\right]+\frac{\lambda}{2(4\pi)^{\frac{D}{2}}}\sum_{j=0}^{\left[\frac{D-1}{2}\right]}A^{(i)}_{\frac{D-1}{2}-j}\Gamma\left(-j+\frac{1}{2}\right)m^{2j-1}\nonumber\\
&+&\frac{\lambda}{2(4\pi)^{\frac{D}{2}}}\sum_{j=D}^{\infty}A^{(i)}_{\frac{j}{2}}m^{D-j-2}\Gamma\left(\frac{j-D}{2}+1\right)\;,
\end{eqnarray}
where $H_{j}$ denotes the $j$-th harmonic number and $[x]$ represents the integer part of $x$.
We would like to make a remark at this point. The procedure just outlined to find the needed counter-terms
can be proved to be equivalent to first writing the spectral zeta function of the operator $-\Delta+M^{2}$ in terms
of the ${\mathscr L}^{2}$-trace of the associated heat kernel and then using the obtained integral representation to rewrite
the one-loop effective action in (\ref{22}). By expanding the resulting expression in terms of $\phi$ one finds that (\ref{31}) is indeed the coefficient
of the expansion proportional to $\phi^{2}$.

The desired counter-terms needed to renormalize
the one-loop mass correction are those proportional to the positive powers of the mass in (\ref{31}). This is equivalent to the
requirement that in the classical limit, namely $m\to\infty$, the quantum corrections to the mass must vanish.
The coefficients $A^{(i)}_{k/2}$ of the asymptotic expansion of the heat kernel in (\ref{30}) can be
computed in terms of the spectral zeta function $\zeta_{i}(s,m)$ \cite{gilkey95,kirsten01,seel68-10-288}. In fact, from the general theory of the spectral zeta function
one has for $k=\{0,\ldots,D-1\}$
\begin{equation}\label{31a}
\tilde{A}^{(i)}_{\frac{k}{2}}=(4\pi)^{\frac{D}{2}}\Gamma\left(\frac{D-k}{2}\right)\textrm{Res}\;\zeta_{i}\left(\frac{D-k}{2},m\right)\;,
\end{equation}
and for $n\in\mathbb{N}_{0}$,
\begin{equation}\label{31b}
\tilde{A}^{(i)}_{\frac{D}{2}-n}=(4\pi)^{\frac{D}{2}}\frac{(-1)^{n}}{n!}\zeta_{i}(-n,m)\;,
\end{equation}
where $\tilde{A}^{(i)}_{j/2}$ represent the \emph{massive} heat kernel coefficients. The relation between the massive
heat kernel coefficients $\tilde{A}^{(i)}_{j/2}$ and the massless ones, namely $A^{(i)}_{j/2}$, appearing in (\ref{30}), can be found to be the following
\begin{equation}\label{31c}
\tilde{A}^{(i)}_{\frac{j}{2}}=\sum_{l=0}^{\left[\frac{j}{2}\right]}\frac{(-1)^{l}}{l!}m^{2l}A^{(i)}_{\frac{j}{2}-l}.
\end{equation}

By subtracting the terms with positive powers of the mass in (\ref{31}) form the right-hand-side of (\ref{28}) and by noticing that
\begin{equation}\label{32}
(4\pi)^{\frac{D}{2}}\textrm{Res}\;\zeta_{i}(1,m)=\sum_{l=0}^{\left[\frac{D}{2}-1\right]}\frac{(-1)^{l}}{l!}A^{(i)}_{\frac{D}{2}-l-1}m^{2l}\;,
\end{equation}
which can be obtained from (\ref{31a}) and (\ref{31c}),
we finally arrive at the following expression for the renormalized one-loop correction to the mass
\begin{eqnarray}\label{33}
\delta m^{2}_{i,\textrm{ren}}&=&-\frac{\lambda}{2(4\pi)^{\frac{D}{2}}V_{d}}\left[\sum_{l=0}^{\left[\frac{D}{2}-1\right]}\frac{(-1)^{l}}{l!}A^{(i)}_{\frac{D}{2}-l-1}m^{2l}\left(H_{l}-\ln m^{2}\right)+\sum_{l=1}^{\left[\frac{D-1}{2}\right]}A^{(i)}_{\frac{D-1}{2}-l}m^{2l-1}\Gamma\left(-l+\frac{1}{2}\right)\right]\nonumber\\
&+&\frac{\lambda}{2V_{d}}\textrm{FP}\;\zeta_{i}(1,m)\;.
\end{eqnarray}
At this point we use (\ref{10}) in (\ref{33}) and express $\delta m^{2}_{i,\textrm{ren}}$ in terms of the functions ${\cal Z}_{i}(s,M|L_{1},\ldots,L_{d})$ introduced
in the previous section. In more detail, by making the replacement $M^{2}\to m^{2}$ in ${\cal Z}_{i}(s,M|L_{1},\ldots,L_{d})$, we get
\begin{eqnarray}\label{34}
\delta m^{2}_{i,\textrm{ren}}&=&-\frac{\lambda}{2(4\pi)^{\frac{D}{2}}V_{D}}\left[\sum_{l=0}^{\left[\frac{D}{2}-1\right]}\frac{(-1)^{l}}{l!}A^{(i)}_{\frac{D}{2}-l-1}m^{2l}\left(H_{l}-\ln m^{2}\right)+\sum_{l=1}^{\left[\frac{D-1}{2}\right]}A^{(i)}_{\frac{D-1}{2}-l}m^{2l-1}\Gamma\left(-l+\frac{1}{2}\right)\right]\nonumber\\
&+&\frac{\lambda}{2V_{D}}\textrm{FP}\;{\cal Z}_{i}(1,m|L_{1},\ldots,L_{D})\;,
\end{eqnarray}
when $D=d$. When, instead, $D-d=2n$, with $n\in\mathbb{N}^{+}$, we have
\begin{eqnarray}\label{35}
\delta m^{2}_{i,\textrm{ren}}&=&-\frac{\lambda}{2(4\pi)^{\frac{D}{2}}V_{d}}\left[\sum_{l=0}^{\left[\frac{D}{2}-1\right]}\frac{(-1)^{l}}{l!}A^{(i)}_{\frac{D}{2}-l-1}m^{2l}\left(H_{l}-\ln m^{2}\right)+\sum_{l=1}^{\left[\frac{D-1}{2}\right]}A^{(i)}_{\frac{D-1}{2}-l}m^{2l-1}\Gamma\left(-l+\frac{1}{2}\right)\right]\nonumber\\
&+&\frac{(-1)^{n-1}\lambda}{2(4\pi)^{n} (n-1)!V_{d}}\left[H_{n-1}{\cal Z}_{i}(1-n,m|L_{1},\ldots,L_{d})+{\cal Z}'_{i}(1-n,m|L_{1},\ldots,L_{d})\right]\;.
\end{eqnarray}
Finally, when $D-d=2n+1$, with $n\in\mathbb{N}_{0}$, we get
\begin{eqnarray}\label{36}
\delta m^{2}_{i,\textrm{ren}}&=&-\frac{\lambda}{2(4\pi)^{\frac{D}{2}}V_{d}}\left[\sum_{l=0}^{\left[\frac{D}{2}-1\right]}\frac{(-1)^{l}}{l!}A^{(i)}_{\frac{D}{2}-l-1}m^{2l}\left(H_{l}-\ln m^{2}\right)+\sum_{l=1}^{\left[\frac{D-1}{2}\right]}A^{(i)}_{\frac{D-1}{2}-l}m^{2l-1}\Gamma\left(-l+\frac{1}{2}\right)\right]\nonumber\\
&+&\frac{\lambda}{2(4\pi)^{n+\frac{1}{2}} V_{d}}\Gamma\left(\frac{1}{2}-n\right)\Bigg\{\textrm{FP}\;{\cal Z}_{i}\left(\frac{1}{2}-n,m|L_{1},\ldots,L_{d}\right)\nonumber\\
&+&\left[\Psi\left(\frac{1}{2}-n\right)+\gamma\right]
\textrm{Res}\;{\cal Z}_{i}\left(\frac{1}{2}-n,m|L_{1},\ldots,L_{d}\right)\Bigg\}\;,
\end{eqnarray}
with $\gamma$ denoting the Euler-Mascheroni constant. To obtain somewhat more explicit expressions for $\delta m^{2}_{i,\textrm{ren}}$ from the formulas (\ref{34})-(\ref{36})
when either Dirichlet, Neumann, or periodic boundary conditions are considered,
we need to use the relations (\ref{19}), (\ref{20}), and (\ref{21}) together with the analytically continued form of the Epstein zeta function. We would like to point
out that it is sufficient to compute $\textrm{FP}\;\zeta_{E}(1)$, $\zeta_{E}(1-n)$, $\zeta'_{E}(1-n)$, $\textrm{Res}\;\zeta_{E}(1/2-n)$,
and $\textrm{FP}\;\zeta_{E}(1/2-n)$ for all the boundary conditions considered in this work since
the functions ${\cal Z}_{D}$, ${\cal Z}_{N}$, and ${\cal Z}_{P}$ are written as linear combinations of the Epstein zeta function.

\section{Derivation of explicit expressions for $\delta m^{2}_{i,ren}$}

In order to compute either the residue, finite part or the values of the Epstein zeta function and its derivative at specific values of $s$ we need its analytically continued expression.
An analytic continuation suitable for our purposes can be obtained by rewriting (\ref{13}) in terms of an integral by using the Mellin transform and by
subsequently employing the Poisson summation formula to obtain \cite{elizalde94a,terras}
\begin{equation}\label{37}
\zeta_{E}\left(s,m^{2}\Bigg|\frac{\pi^2}{L^{2}_{1}},\ldots,\frac{\pi^2}{L^{2}_{d}}\right)=
\frac{V_{d}\Gamma\left(s-\frac{d}{2}\right)}{\pi^{\frac{d}{2}}\Gamma(s)}m^{d-2s}+\frac{m^{\frac{d}{2}-s}V_{d}}{\Gamma(s)}F(s, m |L_{1},\ldots,L_{d})\;,
\end{equation}
where we have defined, for convenience, the function of $s\in\mathbb{C}$
\begin{equation}\label{38}
F(s,m|L_{1},\ldots,L_{d})=2\pi^{-\frac{d}{2}}\sum_{{\bf n}\in\mathbb{Z}^{d}/\{{\bf 0}\}}
\left(\sqrt{L^{2}_{1}n^{2}_{1}+\cdots+L^{2}_{d}n^{2}_{d}}\right)^{s-\frac{d}{2}}K_{s-\frac{d}{2}}\left(2m\sqrt{L^{2}_{1}n^{2}_{1}+\cdots+L^{2}_{d}n^{2}_{d}}\right)\;,
\end{equation}
with $K_{a}(z)$ representing the modified Bessel function of the second kind. Let us point out that in order to obtain the Epstein zeta function
that appears in the case of periodic boundary conditions it is sufficient to perform the replacement $L_{i}\to L_{i}/2$ in (\ref{37}) and (\ref{38}). Due to the
exponentially damped behavior of the modified Bessel function of the second kind, the function $F(s,m|L_1,...,L_d)$ is analytic for $s\in\mathbb{C}$. This implies that
the meromorphic structure of the Epstein zeta function in (\ref{37}) is completely determined by the first term on the right-hand-side of (\ref{37}).

We can use the analytic continuation (\ref{37}) to explicitly compute the needed terms $\textrm{FP}\;\zeta_{E}(1)$, $\zeta_{E}(1-n)$, $\zeta'_{E}(1-n)$, $\textrm{Res}\;\zeta_{E}(1/2-n)$, and $\textrm{FP}\;\zeta_{E}(1/2-n)$.
For the first term in the list we obtain, when $d=2k$, $k\in\mathbb{N}^{+}$,
\begin{equation}\label{39}
\textrm{FP}\;\zeta_{E}\left(1,m^{2}\Bigg|\frac{\pi^2}{L^{2}_{1}},\ldots,\frac{\pi^2}{L^{2}_{2k}}\right)=\frac{(-1)^{k-1}V_{2k}}{(k-1)!\pi^{k}}m^{2k-2}\left(H_{k-1}-\ln m^{2}\right)+m^{k-1}V_{2k}
F(1,m|L_{1},\ldots,L_{2k})\;,
\end{equation}
and, when $d=2k+1$, for $k\in\mathbb{N}_{0}$,
\begin{equation}\label{40}
\textrm{FP}\;\zeta_{E}\left(1,m^{2}\Bigg|\frac{\pi^2}{L^{2}_{1}},\ldots,\frac{\pi^2}{L^{2}_{2k+1}}\right)=
\frac{V_{2k+1}}{\pi^{k+\frac{1}{2}}}m^{2k-1}\Gamma\left(-k+\frac{1}{2}\right)+m^{k-\frac{1}{2}}V_{2k+1}
F(1,m|L_{1},\ldots,L_{2k+1})\;.
\end{equation}
For the next term, for even and odd values of the dimension $d$, namely $d=2k$, $k\in\mathbb{N}^{+}$, and $d=2k+1$, with $k\in\mathbb{N}_{0}$, we have
\begin{equation}\label{41}
\zeta_{E}\left(1-n,m^{2}\Bigg|\frac{\pi^2}{L^{2}_{1}},\ldots,\frac{\pi^2}{L^{2}_{2k}}\right)=\frac{(-1)^{k}(n-1)!V_{2k}}{(n+k-1)!\pi^{k}}m^{2k+2n-2}\;,\quad\textrm{and}\quad
\zeta_{E}\left(1-n,m^{2}\Bigg|\frac{\pi^2}{L^{2}_{1}},\ldots,\frac{\pi^2}{L^{2}_{2k+1}}\right)=0\;,
\end{equation}
where $n\in\mathbb{N}^{+}$.
The derivative of the Epstein zeta function at negative integers reads, for even values of $d$,
\begin{eqnarray}\label{42}
\zeta'_{E}\left(1-n,m^{2}\Bigg|\frac{\pi^2}{L^{2}_{1}},\ldots,\frac{\pi^2}{L^{2}_{2k}}\right)&=&\frac{(-1)^{k}(n-1)!V_{2k}}{(n+k-1)!\pi^{k}}m^{2k+2n-2}
\left(H_{n+k-1}-H_{n-1}-\ln m^{2}\right)\nonumber\\
&+&(-1)^{n-1}(n-1)!\,m^{k+n-1}V_{2k}
F(1-n,m|L_{1},\ldots,L_{2k})\;,
\end{eqnarray}
and, for odd values of the dimension $d$, one has
\begin{eqnarray}\label{43}
\zeta'_{E}\left(1-n,m^{2}\Bigg|\frac{\pi^2}{L^{2}_{1}},\ldots,\frac{\pi^2}{L^{2}_{2k+1}}\right)&=&\frac{(-1)^{n-1}(n-1)!V_{2k+1}}{\pi^{k+\frac{1}{2}}}m^{2k+2n-1}\Gamma\left(-n-k+\frac{1}{2}\right)\nonumber\\
&+&(-1)^{n}(n-1)!\,m^{k+n-\frac{1}{2}}V_{2k+1}
F(1-n,m|L_{1},\ldots,L_{2k+1})\;.
\end{eqnarray}
Lastly, the finite part and the residue of the Epstein zeta function at negative half-integers are, for $d=2k$, $k\in\mathbb{N}^{+}$,
\begin{eqnarray}\label{44}
\textrm{FP}\;\zeta_{E}\left(\frac{1}{2}-n,m^{2}\Bigg|\frac{\pi^2}{L^{2}_{1}},\ldots,\frac{\pi^2}{L^{2}_{2k}}\right)&=&\frac{V_{2k}}{\pi^{k}}m^{2k+2n-1}
\frac{\Gamma\left(-n-k+\frac{1}{2}\right)}{\Gamma\left(\frac{1}{2}-n\right)}+\frac{m^{n+k-\frac{1}{2}}V_{2k}}{\Gamma\left(\frac{1}{2}-n\right)}
F\left(\frac{1}{2}-n,m \Bigg|L_{1},\ldots,L_{2k}\right)\;,\nonumber\\
\textrm{Res}\;\zeta_{E}\left(\frac{1}{2}-n,m^{2}\Bigg|\frac{\pi^2}{L^{2}_{1}},\ldots,\frac{\pi^2}{L^{2}_{2k}}\right)&=&0 ,
\end{eqnarray}
and, for $d=2k+1$, with $k\in\mathbb{N}_{0}$, we obtain
\begin{eqnarray}\label{45}
\textrm{FP}\;\zeta_{E}\left(\frac{1}{2}-n,m^{2}\Bigg|\frac{\pi^2}{L^{2}_{1}},\ldots,\frac{\pi^2}{L^{2}_{2k+1}}\right)&=&\frac{(-1)^{n+k}V_{2k+1}}{(n+k)!\pi^{k+\frac{1}{2}}
\Gamma\left(\frac{1}{2}-n\right)}m^{2k+2n}\left(H_{n+k}-\ln m^{2}+2\ln 2-2\sum_{j=1}^{n}\frac{1}{2j-1}\right)\nonumber\\
&+&\frac{m^{n+k}V_{2k+1}}{\Gamma\left(\frac{1}{2}-n\right)}
F\left(\frac{1}{2}-n,m \Bigg|L_{1},\ldots,L_{2k+1}\right)\;,\nonumber\\
\textrm{Res}\;\zeta_{E}\left(\frac{1}{2}-n,m^{2}\Bigg|\frac{\pi^2}{L^{2}_{1}},\ldots,\frac{\pi^2}{L^{2}_{2k+1}}\right)&=&\frac{(-1)^{n+k}V_{2k+1}}{(n+k)!\pi^{k+\frac{1}{2}}\Gamma\left(\frac{1}{2}-n\right)}m^{2k+2n}\;.
\end{eqnarray}

The results that we have obtained for the Epstein zeta function in (\ref{39}) through (\ref{45}) together with the relations
\begin{eqnarray}\label{46}
{\cal Z}_{D}(s,m|L_{1},\ldots,L_{d})&=&\frac{(-1)^{d}}{2^{d}}m^{-2s}+\frac{(-1)^{d}}{2^{d}}\sum_{l=1}^{\left[\frac{d}{2}\right]}\sum_{1\leq i_{1}<i_{2}<\cdots<i_{2l}\leq d}\zeta_{E}\left(s,m^{2}\Bigg|\frac{\pi^{2}}{L^{2}_{i_{1}}},\ldots,\frac{\pi^{2}}{L^{2}_{i_{2l}}}\right)\nonumber\\
&-&\frac{(-1)^{d}}{2^{d}}\sum_{l=0}^{\left[\frac{d-1}{2}\right]}\sum_{1\leq i_{1}<i_{2}<\cdots<i_{2l+1}\leq d}\zeta_{E}\left(s,m^{2}\Bigg|\frac{\pi^{2}}{L^{2}_{i_{1}}},\ldots,\frac{\pi^{2}}{L^{2}_{i_{2l+1}}}\right)\;,
\end{eqnarray}
and
\begin{eqnarray}\label{47}
{\cal Z}_{N}(s,m|L_{1},\ldots,L_{d})&=&m^{-2s}+\sum_{l=1}^{\left[\frac{d}{2}\right]}\sum_{1\leq i_{1}<i_{2}<\cdots<i_{2l}\leq d}\zeta_{E}\left(s,m^{2}\Bigg|\frac{\pi^{2}}{L^{2}_{i_{1}}},\ldots,\frac{\pi^{2}}{L^{2}_{i_{2l}}}\right)\nonumber\\
&+&\sum_{l=0}^{\left[\frac{d-1}{2}\right]}\sum_{1\leq i_{1}<i_{2}<\cdots<i_{2l+1}\leq d}\zeta_{E}\left(s,m^{2}\Bigg|\frac{\pi^{2}}{L^{2}_{i_{1}}},\ldots,\frac{\pi^{2}}{L^{2}_{i_{2l+1}}}\right)\;,
\end{eqnarray}
which can be easily derived from (\ref{19}) and (\ref{20}), respectively, can be used to evaluate the needed terms involving ${\cal Z}_{i}$ in the
expressions for the one-loop mass correction (\ref{34})-(\ref{36}). Obviously the formulas (\ref{39}) through (\ref{45}) with $L_{i}\to L_{i}/2$ are sufficient for
calculating $\delta m^{2}_{P,\textrm{ren}}$.

In addition to the results just found above for the needed terms
involving the functions ${\cal Z}_{i}$, we also need to compute the heat kernel coefficients that appear in (\ref{34})-(\ref{36}). This will finally provide more explicit expressions for the one-loop mass correction $\delta m^{2}_{P,\textrm{ren}}$. According to (\ref{34})-(\ref{36}) we only need the heat kernel coefficients up to and including $A_{D/2-1}$.
These coefficients can be obtained as follows: First, the massive heat kernel coefficients with $k=\{0,\ldots, D-2\}$ are computed by exploiting (\ref{31a}) and (\ref{10}) through the formula
\begin{equation}\label{48}
\tilde{A}^{(i)}_{\frac{k}{2}}=(4\pi)^{\frac{d}{2}}\textrm{Res}\;\left[\Gamma\left(\frac{d-k}{2}\right){\cal Z}_{i}\left(\frac{d-k}{2},m\Bigg|L_{1},\ldots,L_{d}\right)\right]\;,
\end{equation}
and then the massless ones are derived by using (\ref{48}) in the relation (\ref{31c}).
We can further evaluate the residue contained in the expression (\ref{48}). For $k=\{0,\ldots,d-1\}$ we have
\begin{equation}\label{49}
\tilde{A}^{(i)}_{\frac{k}{2}}=(4\pi)^{\frac{d}{2}}\Gamma\left(\frac{d-k}{2}\right)\textrm{Res}\;{\cal Z}_{i}\left(\frac{d-k}{2},m\Bigg|L_{1},\ldots,L_{d}\right)\;,
\end{equation}
while when $k=d+2j$, with $j=\{0,\ldots,[(D-d)/2-1]\}$,
\begin{equation}\label{50}
\tilde{A}^{(i)}_{\frac{d}{2}+j}=(4\pi)^{\frac{d}{2}}\frac{(-1)^{j}}{j!}{\cal Z}_{i}\left(-j,m|L_{1},\ldots,L_{d}\right)\;,
\end{equation}
and finally when $k=d+2j+1$, with $j=\{0,\ldots,[(D-d-3)/2]\}$, one gets
\begin{equation}\label{51}
\tilde{A}^{(i)}_{\frac{d+1}{2}+j}=(4\pi)^{\frac{d}{2}}\Gamma\left(-j-\frac{1}{2}\right)\textrm{Res}\;{\cal Z}_{i}\left(-j-\frac{1}{2},m\Bigg|L_{1},\ldots,L_{d}\right)\;.
\end{equation}
As we have already mentioned earlier, the functions ${\cal Z}_{i}$ are expressed in terms of a linear combination of the Epstein zeta function.
This implies that in order to compute the heat kernel coefficients (\ref{49}) through (\ref{51}) it is sufficient to consider the residues and
the value at negative integers of the Epstein zeta function. From the analytically continued expression for the Epstein zeta function in (\ref{37})
one can prove that \cite{elizalde95,kirsten10} for even values of the dimension $d$ the Epstein zeta function $\zeta_{E}(s,m^{2}|{\bf r})$ has simple poles at the points
$s=(d-k)/2$, with $k=\{0,\ldots,d-2\}$, while for odd values of $d$ the simple poles are located at the points $s=(d-k)/2$, with $k=\{0,\ldots,d-1\}$, and
at the negative half-integer points $s=-(2l+1)/2$, with $l\in\mathbb{N}_{0}$. The residues are found to be
\begin{equation}\label{52}
\textrm{Res}\;\zeta_{E}\left(j,m^{2}|{\bf r}\right)=\frac{(-1)^{\frac{d}{2}-j}\pi^{\frac{d}{2}}m^{d-2j}}{\sqrt{r_{1}\cdots r_{d}}\,\Gamma(j)\Gamma\left(\frac{d}{2}-j+1\right)}\;,
\end{equation}
whereas for the values of $\zeta_{E}(s,m^{2}|{\bf r})$ at the negative integers we have
\begin{equation}\label{53}
\zeta_{E}\left(-p,m^{2}|{\bf r}\right)=\frac{(-1)^{\frac{d}{2}}p!\pi^{\frac{d}{2}}m^{d+2p}}{\sqrt{r_{1}\cdots r_{d}}\,\Gamma\left(\frac{d}{2}+j+1\right)}\;,\quad\textrm{for $d$ even},\quad \zeta_{E}\left(-p,m^{2}|{\bf r}\right)=0\;,\quad\textrm{for $d$ odd}\;.
\end{equation}

The massive heat kernel coefficients for the case of periodic boundary conditions can be found from (\ref{49})-(\ref{51}) by exploiting the definition
(\ref{21}) and the results (\ref{52}) and (\ref{53}). In more detail we obtain, for $j=\{0,\ldots,[D/2-1]\}$,
\begin{equation}\label{54}
\tilde{A}^{(P)}_{j}=\frac{(-1)^{j}m^{2j}}{j!}V_{d}\;.
\end{equation}
Obviously, the massless coefficients can be computed from the massive ones by setting $m=0$ to get
\begin{equation}\label{55}
A^{(P)}_{j}=
  \begin{cases}
    \;\;V_{d} & \text{if $j=0$}\;,\\
    \;\;0 & \text{otherwise}\;.
  \end{cases}
\end{equation}
The above result was to be expected since, for periodic boundary conditions, the space reduces to a higher-dimensional torus and, for this geometry,
the only non-vanishing heat kernel coefficient is the first one which corresponds to the volume of the torus.
The coefficients $\tilde{A}^{(D)}_{k/2}$ for Dirichlet boundary conditions can be computed by exploiting (\ref{46}) and the relations (\ref{49})
through (\ref{53}). After a lengthy, yet straightforward, calculation one obtains for $k=\{0,\ldots,D-2\}$
\begin{equation}\label{56}
\tilde{A}^{(D)}_{\frac{k}{2}}=\sum_{n=\textrm{max}\{[(k-d+1)/2],0\}}^{\left[\frac{k}{2}\right]}\frac{(-1)^{n+k}m^{2n}}{\pi^{n-\frac{k}{2}}n!}
\sum_{1\leq i_{1}<\cdots<i_{2n+d-k}\leq d}L_{i_{1}}\cdots L_{i_{2n+d-k}}\;.
\end{equation}
By comparing (\ref{46}) and (\ref{47}) and by keeping in mind the formulas (\ref{49})-(\ref{51}) it is not difficult to realize that the
massive heat kernel coefficients for the Neumann case can be obtained from the corresponding coefficients for Dirichlet boundary conditions
as follows
\begin{equation}\label{57}
\tilde{A}^{(N)}_{\frac{k}{2}}=(-1)^{k}\tilde{A}^{(D)}_{\frac{k}{2}}\;,
\end{equation}
for $k=\{0,\ldots,D-2\}$.
Once again, in order to obtain the massless heat kernel coefficients it is sufficient to set $m=0$ in the expression (\ref{56}). In more detail we find
\begin{equation}\label{58}
A^{(D)}_{\frac{k}{2}}=
  \begin{cases}
    \;\; (-1)^{k}\pi^{\frac{k}{2}}\sum_{1\leq i_{1}<\cdots<i_{d-k}\leq d}L_{i_{1}}\cdots L_{i_{d-k}}&\text{when $k=\{0,\ldots, d-1\}$}\;,\\
    \;\; (-1)^{d}\pi^{\frac{d}{2}}&\text{when $k=d$}\;,\\
    \;\;0 & \text{when $k\geq d$}\;,
  \end{cases}
\end{equation}
for Dirichlet boundary conditions and, according to (\ref{57}), we have
\begin{equation}\label{59}
A^{(N)}_{\frac{k}{2}}=
  \begin{cases}
    \;\; \pi^{\frac{k}{2}}\sum_{1\leq i_{1}<\cdots<i_{d-k}\leq d}L_{i_{1}}\cdots L_{i_{d-k}}&\text{when $k=\{0,\ldots, d-1\}$}\;,\\
    \;\; \pi^{\frac{d}{2}}&\text{when $k=d$}\;,\\
    \;\;0 & \text{when $k\geq d$}\;,
  \end{cases}
\end{equation}
for Neumann boundary conditions.

We are finally in the position to compute explicit expressions for the one-loop mass corrections $\delta m^{2}_{i,\textrm{ren}}$.
For periodic boundary conditions, we use (\ref{34})-(\ref{36}) and the results presented in this section to obtain, after a somewhat protracted
calculation, for all values of the dimension $d$ and $D$
\begin{equation}\label{60}
\delta m^{2}_{P,\textrm{ren}}=\frac{\lambda}{2^{D+1}\pi^{\frac{D-d}{2}}}m^{\frac{D}{2}-1}
F\left(1-\frac{D-d}{2},m \Bigg|\frac{L_{1}}{2},\ldots ,\frac{L_{d}}{2}\right)\;.
\end{equation}
For the case of Dirichlet boundary conditions a similar calculation leads to the following result, for $D\geq d$,
\begin{eqnarray}\label{62}
\delta m^{2}_{D,\textrm{ren}}&=&\frac{(-1)^{d}\lambda m^{\frac{D-d}{2}}}{2^{d+1}(4\pi)^{\frac{D-d}{2}}V_{d}}
\Bigg[\delta_{D,d}\Bigg(\frac{1}{m^{2}}-\frac{1}{m}\sum_{i=1}^{d}L_{i}\Bigg)+\frac{\delta_{D,d+1}\sqrt{\pi}}{m^{\frac{3}{2}}}\nonumber\\
&+&\sum_{l=1}^{\left[\frac{d}{2}\right]}m^{l-1}\sum_{1\leq i_{1}<\cdots<i_{2l}\leq d}L_{i_{1}}\cdots L_{i_{2l}}
F\left(1-\frac{D-d}{2},m \Bigg| L_{i_{1}},\ldots L_{i_{2l}}\right)\nonumber\\
&-&\sum_{l=0}^{\left[\frac{d-1}{2}\right]}m^{l-\frac{1}{2}}\sum_{1\leq i_{1}<\cdots<i_{2l+1}\leq d}L_{i_{1}}\cdots L_{i_{2l+1}}
F\left(1-\frac{D-d}{2},m \Bigg| L_{i_{1}},\ldots L_{i_{2l+1}}\right)\Bigg]\;,
\end{eqnarray}
where $\delta_{i,j}$ denotes the Kronecker delta function. Similarly, for Neumann boundary conditions we obtain, for $D\geq d$,
\begin{eqnarray}\label{64}
\delta m^{2}_{N,\textrm{ren}}&=&\frac{\lambda m^{\frac{D-d}{2}}}{2^{d+1}(4\pi)^{\frac{D-d}{2}}V_{d}}
\Bigg[\delta_{D,d}\Bigg(\frac{1}{m^{2}}+\frac{1}{m}\sum_{i=1}^{d}L_{i}\Bigg)+\frac{\delta_{D,d+1}\sqrt{\pi}}{m^{\frac{3}{2}}}\nonumber\\
&+&\sum_{l=1}^{\left[\frac{d}{2}\right]}m^{l-1}\sum_{1\leq i_{1}<\cdots<i_{2l}\leq d}L_{i_{1}}\cdots L_{i_{2l}}
F\left(1-\frac{D-d}{2},m \Bigg| L_{i_{1}},\ldots L_{i_{2l}}\right)\nonumber\\
&+&\sum_{l=0}^{\left[\frac{d-1}{2}\right]}m^{l-\frac{1}{2}}\sum_{1\leq i_{1}<\cdots<i_{2l+1}\leq d}L_{i_{1}}\cdots L_{i_{2l+1}}
F\left(1-\frac{D-d}{2},m \Bigg| L_{i_{1}},\ldots L_{i_{2l+1}}\right)\Bigg]\;.
\end{eqnarray}

The above results for the one-loop correction to the mass of the $\lambda\phi^{4}$ theory  allows us to write the \emph{renormalized}
mass of the theory as follows
\begin{equation}\label{65a}
m^{2}_{i,\textrm{{\it ren}}}=m^{2}+\delta m_{i,\textrm{{\it ren}}}^{2}\;.
\end{equation}
It is clear, from the results (\ref{60}) through (\ref{64}) that the renormalized mass depends explicitly
on the geometry of the space and the boundary conditions imposed.

\section{Application of the one-loop mass correction to the Ginzburg-Landau model}

In the Ginzburg-Landau theory the Hamiltonian density describing the dynamics of a complex order parameter $\Psi$ in a $D$-dimensional Euclidean space endowed with a
$d$-dimensional compactified subspace is the following
\begin{equation}\label{66}
{\cal H}=\frac{1}{2}|\nabla\Psi|^{2}+\frac{a^{2}(L_{i})}{2}|\Psi|^{2}+\frac{\lambda}{4!}|\Psi|^{4}\;,
\end{equation}
where $\lambda$ is the renormalized coupling constant and $a^{2}$ is a mass parameter that depends on the compactification lengths $L_{i}$ \cite{abreu05}.
The theory defined by the Hamiltonian (\ref{66}) is fundamentally a mean-field theory which was introduced to describe second-order
phase transitions in neutral superconductors \cite{abreu05,annett}.
In the Ginzburg-Landau theory on the unbounded $D$-dimensional Euclidean space, the mass parameter is related, in the vicinity of criticality,
to the critical temperature $T_{c}$ as follows
\begin{equation}\label{77}
a^{2}\simeq \alpha (T-T_{c})\;,
\end{equation}
where $\alpha>0$ is a constant independent of the temperature. When a compactified $d$-dimensional subspace is introduced, the mass parameter in (\ref{66})
depends on the lengths $L_{i}$, and consequently, on the boundary conditions imposed. In this case the $L_{i}$-dependent mass parameter defines
an associated $L_{i}$-dependent critical temperature as follows
\begin{equation}\label{78}
a^{2}(L_{i})\simeq \alpha (T-T_{c}(L_{i}))\;.
\end{equation}
The Euclidean critical temperature $T_{c}$ can be recovered from the boundary modified critical temperature $T_{c}(L_{i})$ through the
limit
\begin{equation}\label{79}
\lim_{(L_{1},\ldots, L_{d})\to\infty}T_{c}(L_{1},\ldots, L_{d})=T_{c}\;.
\end{equation}

By comparing (\ref{66}) with (\ref{1}) it is easy to realize that the Ginzburg-Landau theory
for a neutral superconductor is equivalent to an Euclidean self-interacting scalar field theory. This implies that the methods we employed to
study the fluctuations of the self-interacting scalar field are appropriate to analyze the fluctuations of the order parameter $\Psi$. In particular
we can utilize the results obtained in the previous sections for the one-loop mass correction to analyze how the presence of
a compact subspace modifies the factor $a^{2}$ and, consequently, the critical temperature.
For the sake of simplicity we analyze first the case of periodic boundary conditions, namely we consider the compact subspace
to have the topology of a $d$-dimensional torus. This is also one of the most widely studied configurations in the literature (see e.g. \cite{khanna14} and
references therein).

In the framework of Ginzburg-Landau theory, the equation for the mass parameter is obtained, in the neighborhood of criticality, from the length-dependent
gap equation \cite{malbo03,khanna09,khanna14}. In this limit, the length-dependent gap equation reduces to a Dyson-Schwinger type equation for the mass parameter
which has the same form as the eq. (\ref{65a}), namely \cite{malbo02}
\begin{equation}\label{80}
a_{P}^{2}(L_{i})=a^{2}+\delta a_{P}^{2}(L_{i})\;,
\end{equation}
where $\delta a_{P}^{2}$ is given by (\ref{60}) once one performs the replacement $m\to a_{P}(L_{i})$. Near the
critical temperature we hence have, for any dimensions $D$ and $d$ with $D\geq d$, the expression \cite{malbo02}
\begin{equation}\label{81}
a_{P}^{2}(L_{i})=a^{2}+\frac{\lambda a_{P}^{\frac{D}{2}-1}(L_{i})}{2^{\frac{D}{2}+1}\pi^{\frac{D}{2}}}\sum_{{\bf n}\in\mathbb{Z}^{d}/\{{\bf 0}\}}
\left(\sqrt{L^{2}_{1}n^{2}_{1}+\cdots+L^{2}_{d}n^{2}_{d}}\right)^{1-\frac{D}{2}}K_{1-\frac{D}{2}}\left(a_{P}(L_{i})\sqrt{L^{2}_{1}n^{2}_{1}+\cdots+L^{2}_{d}n^{2}_{d}}\right)\;.
\end{equation}
Eq. (\ref{81}) cannot be solved for $a_{P}(L_{i})$, and hence for the $L_{i}$-dependent critical temperature, in general, however one can attempt a solution in the
neighborhood of criticality by expanding (\ref{81}) for small $a_{P}(L_{i})$. The expansion of the series appearing in (\ref{81}) can be performed
by following the method outlined in \cite{fucci15}. The method requires rewriting the series in (\ref{81}) by using the complex integral representation
of the modified Bessel function of the second kind. In detail, in Section 3 of \cite{fucci15} we considered
\begin{eqnarray}\label{82}
g(s,q|L_{1},\ldots,L_{d})&=&q^{s}\sum_{{\bf n}\in\mathbb{Z}^{d}/\{{\bf 0}\}}
\left(\sqrt{L^{2}_{1}n^{2}_{1}+\cdots+L^{2}_{d}n^{2}_{d}}\right)^{-s}K_{-s}\left(2q\sqrt{L^{2}_{1}n^{2}_{1}+\cdots+L^{2}_{d}n^{2}_{d}}\right)\nonumber\\
&=&\frac{1}{4\pi i}\int_{c-i\infty}^{c+i\infty}\Gamma(t)\Gamma(t+s)q^{-2t}\zeta_{E}(s+t|L_{1},\ldots, L_{d})\diff t\;,
\end{eqnarray}
where $c>\textrm{max}\{0,d/2-\Re(s)\}$ and $\zeta_{E}(u|L_{1},\ldots, L_{d})$ denotes the homogeneous Epstein zeta function
\begin{equation}
\zeta_{E}(u|L_{1},\ldots, L_{d})=\sum_{{\bf n}\in\mathbb{Z}^{d}/\{{\bf 0}\}}\left(L^{2}_{1}n_{1}^{2}+L_{2}^{2}n_{2}^{2}+\ldots +L^{2}_{d}n_{d}^{2}\right)^{-u}\;.
\end{equation}
By closing the integration contour to the left and by noticing that $\zeta_{E}(u|L_{1},\ldots, L_{d})$ has a single simple pole
at $s=d/2$ having residue \cite{elizalde95}
\begin{equation}
\textrm{Res}\;\zeta_{E} \left(\frac{d}{2}\Bigg|L_{1},\ldots, L_{d}\right)=\frac{\pi^{d/2}}{\Gamma(d/2)L_{1}\cdots L_{d}}\;,
\end{equation}
and that $\zeta_{E}(-n|L_{1},\ldots, L_{d})=0$, with $n\in\mathbb{N}^{+}$, and also that $\zeta_{E}(0|L_{1},\ldots, L_{d})=-1$,
one can use the residue theorem to obtain the desired small-$q$ asymptotic expansion. Since we are mainly interested
in the small-$a_P (L_i)$ expansion of (\ref{81}) it is convenient to focus our analysis on the values $s=D/2-1$, namely $s=n$, $n\in\mathbb{N}_{0}$
when $D$ is even and $s=(2n+1)/2$ when $D$ is odd. For these particular cases one finds (cf. \cite{fucci15}) for $d=2l$, $l\in\mathbb{N}^{+}$
\begin{eqnarray}\label{74a}
g(n,q|L_{1},\ldots,L_{2l})&\sim&\frac{1}{2}\sum_{j=0\atop j\neq n-l}^{n-1}\frac{(-1)^{j}}{j!}\Gamma(n-j)q^{2j}\zeta_{E}(n-j|L_{1},\ldots,L_{2l})\nonumber\\
&+&\frac{(-1)^{n}}{2}\sum_{j=n}^{\infty}\frac{q^{2j}}{j!(j-n)!}\zeta_{E}^{\prime}(n-j|L_{1},\ldots,L_{2l})\nonumber\\
&+&\frac{(-1)^{n}}{n!}q^{2n}\left(\gamma+\ln q-2H_{n}\right)+\Theta(l-n-1)\frac{q^{2n-2l}\pi^{l}}{2L_{1}\cdots L_{2l}}\Gamma(l-n)\nonumber\\
&+&\Theta(n-l)\frac{(-1)^{n-l}q^{2n-2l}\pi^{l}}{2(n-l)!L_{1}\cdots L_{2l}}\Big[\pi^{-l}(l-1)!L_{1}\cdots L_{2l}\textrm{FP}\,\zeta_{E}(l|L_{1},\ldots,L_{2l})\nonumber\\
&+&\Psi(n-l+1)+\Psi(l)-2\ln q \Big]\;,
\end{eqnarray}
and
\begin{eqnarray}\label{74b}
g\left(\frac{2n+1}{2},q\Bigg|L_{1},\ldots,L_{2l}\right)&\sim&\frac{1}{2}\sum_{m=0}^{\infty}\frac{(-1)^{m}}{m!}\Gamma\left(n-m+\frac{1}{2}\right)q^{2m}\zeta_{E}\left(n-m+\frac{1}{2}\Bigg|L_{1},\ldots,L_{2l}\right)\nonumber\\
&-&\frac{q^{2n+1}}{2}\Gamma\left(-n-\frac{1}{2}\right)
+\frac{q^{2n-2l+1}\pi^l}{2L_{1}\cdots L_{2l}}\Gamma\left(l-n-\frac{1}{2}\right)\;.
\end{eqnarray}
When, instead, $d=2l+1$, $l\in\mathbb{N}_{0}$, one finds
\begin{eqnarray}\label{74c}
g(n,q|L_{1},\ldots,L_{2l+1})&\sim&\frac{1}{2}\sum_{j=0}^{n-1}\frac{(-1)^{j}}{j!}\Gamma(n-j)q^{2j}\zeta_{E}(n-j|L_{1},\ldots,L_{2l+1})\nonumber\\
&+&\frac{(-1)^{n}}{2}\sum_{j=n}^{\infty}\frac{q^{2j}}{j!(j-n)!}\zeta_{E}^{\prime}(n-j|L_{1},\ldots,L_{2l+1})\nonumber\\
&+&\frac{(-1)^{n}}{n!}q^{2n}\left(\gamma+\ln q -2H_{n}\right)+\frac{\pi^{l+\frac{1}{2}}q^{-2l+2n-1}}{2L_{1}\cdots L_{2l+1}}\Gamma\left(l-n+\frac{1}{2}\right)\;,\nonumber\\
\end{eqnarray}
and
\begin{eqnarray}\label{74e}
\lefteqn{g\left(\frac{2n+1}{2},q \Bigg|L_{1},\ldots,L_{2l+1}\right)\sim\frac{1}{2}\sum_{j=0\atop j\neq n-l}^{\infty}\frac{(-1)^{j}}{j!}\Gamma\left(n-j+\frac{1}{2}\right)q^{2j}\zeta_{E}\left(n-j+\frac{1}{2}\Bigg|L_{1},\ldots,L_{2l+1}\right)}\nonumber\\
&&-\frac{q^{2n+1}}{2}\Gamma\left(-n-\frac{1}{2}\right)\nonumber\\
&+&\Theta(n-l)\frac{(-1)^{n-l}q^{2n-2l}\pi^{l+\frac{1}{2}}}{2(n-l)!L_{1}\cdots L_{2l+1}}\Bigg[\pi^{-l-\frac{1}{2}}\Gamma\left(l+\frac{1}{2}\right)\textrm{FP}\,\zeta_{E}\left(l+\frac{1}{2}\Bigg|L_{1},\ldots,L_{2l+1}\right)L_{1}\cdots L_{2l+1}\nonumber\\
&+&\Psi(n-l+1)+\Psi\left(l+\frac{1}{2}\right)-2\ln q\Bigg]+\Theta(l-n-1)\frac{q^{2n-2l}\pi^{l+\frac{1}{2}}}{2L_{1}\cdots L_{2l+1}}\Gamma(l-n)\;.
\end{eqnarray}
We can now use the general results obtained above to study the physically relevant case of $D=3$ and $d\leq 3$. When $D=3$ and $d=2$, close to criticality, namely
$a_{P}(L_{i})\to 0$, we use (\ref{74b}) with $q\to a_P (L_i)$, $L_i\to L_i/2$, to obtain an expansion for (\ref{81}). To the leading order one has
\begin{equation}\label{83}
a_{P}^{2}(L_{i})\simeq a^{2}+\frac{\lambda}{16 L_{1}L_{2}}a_{P}^{-1}(L_{i})\;.
\end{equation}
By using (\ref{77}) and (\ref{78}), the expression (\ref{83}) represents an implicit equation for the $L_{i}$-dependent critical temperature $T_{c}(L_{i})$.
When $D=3$ and $d=1$, for $a_{P}(L_{i})\to 0$ we need to exploit the expansion (\ref{74e}) with the given replacements. In this case
it is not very difficult to obtain, from (\ref{81}), the relation
\begin{equation}\label{86}
a_{P}^{2}(L)\simeq a^{2}-\frac{\lambda}{4\pi L}\ln \left[ a_{P}(L)\right]\;.
\end{equation}
Lastly, for $D=d=3$ one uses, once again, the result in (\ref{74e}) with suitable replacements to get the expression
\begin{equation}\label{87}
a_{P}^{2}(L_{i})\simeq a^{2}+\frac{\lambda}{32 L_{1}L_{2}L_{3}}a_{P}^{-2}(L_{i})\;,
\end{equation}
which, just like the previous ones, represents an implicit equation for $T_{c}(L_{i})$.
It is important to point out that in the equations obtained in (\ref{83}), (\ref{86}), and (\ref{87}) the $L_{i}$-dependent terms
vanish as $L_i\to\infty$, as one should expect since for $L_{i}\to\infty$, $T_{c}(L_{i})$ must reduce to the Euclidean critical temperature $T_{c}$.

As we have already mentioned above, the case involving a toroidally compactified subspace has been extensively studied in the literature.
Here, we would like to extend the results for the critical temperature to compact subspaces with Dirichlet and Neumann boundary conditions, namely to
a $d$-dimensional box embedded in a $D$-dimensional Euclidean space.
In this case, the $L_{i}$-dependent mass parameter in the Ginzburg-Landau model still satisfies the eq. (\ref{80}) where, for the case
of Dirichlet and Neumann boundary conditions, the one-loop correction to consider is $\delta a_{D}^{2}(L_{i})$ and $\delta a_{N}^{2}(L_{i})$, respectively.
Just like the periodic case, $\delta a_{D}^{2}(L_{i})$ and $\delta a_{N}^{2}(L_{i})$ are obtained from (\ref{62}) and (\ref{64}) with the
replacement $m\to a_{D}(L_{i})$ and $m\to a_{N}(L_{i})$, respectively. We will be focusing, once again, on the physically relevant case of $D=3$.

We consider first Dirichlet boundary conditions. In this case the relation between the $L_{i}$-dependent mass parameter and the Euclidean one is
\begin{equation}\label{87a}
a_{D}^{2}(L_{i})=a^{2}+\delta a_{D}^{2}(L_{i})\;.
\end{equation}

For $D=3$ and $d=2$, we exploit the expression (\ref{62}) with $m\to a_{D}(L_{i})$ and the relation
\begin{equation}\label{87b}
F(s,m|L_{1}\ldots,L_{d})=2\pi^{-\frac{d}{2}}m^{s-\frac{d}{2}}g\left(\frac{d}{2}-s,m\Bigg|L_{1}\ldots,L_{d}\right)\;,
\end{equation}
to obtain
\begin{equation}\label{88}
\delta a_{D}^{2}(L_{i})=\frac{\lambda}{16L_{1}L_{2}}\left[\frac{2L_{1}L_{2}}{\pi^{\frac{3}{2}}}g\left(\frac{1}{2},a_{D}(L_{i})\Bigg|L_{1},L_{2}\right)-\frac{2L_{1}}{\pi}g(0,a_{D}(L_{i})|L_{1})
-\frac{2L_{2}}{\pi}g(0,a_{D}(L_{i})|L_{2})+a_{D}^{-1}(L_{i})\right].
\end{equation}
By using (\ref{74b}) and (\ref{74c}) we obtain, for $a_{D}(L_{i})\to 0$, the relation
\begin{equation}\label{89}
a_{D}^{2}(L_{i})\simeq a^{2}-\frac{\lambda}{8\pi L_{1}L_{2}}(L_{1}+L_{2})\ln[a_{D}(L_{i})]\;.
\end{equation}

For $D=3$ and $d=1$ we use (\ref{62}) and the relation (\ref{87b}) to write
\begin{equation}\label{90}
\delta a_{D}^{2}(L)=-\frac{\lambda}{8\pi\sqrt{\pi}}g\left(\frac{1}{2},a_{D}(L)\Bigg|L\right)\;.
\end{equation}
By using (\ref{90}) in (\ref{87a}), we find, in the limit $a_{D}(L)\to 0$ the expression
\begin{equation}\label{91}
a_{D}^{2}(L)\simeq a^{2}-\frac{\lambda}{16\pi L}\ln \left[a_{D}(L)\right]\;.
\end{equation}

Finally, when $D=d=3$, we utilize (\ref{62}) with the appropriate redefinition of the mass term to get
\begin{eqnarray}\label{92}
a_{D}^{2}(L_{i})&\simeq& -\frac{\lambda}{16L_{1}L_{2}L_{3}}\Bigg[a_{D}^{-2}(L_{i})+\frac{2L_{1}L_{2}}{\pi}g(0,a_{D}(L_{i})|L_{1},L_{2})+\frac{2L_{1}L_{3}}{\pi}g(0,a_{D}(L_{i})|L_{1},L_{3})\nonumber\\
&+&\frac{2L_{2}L_{3}}{\pi}g(0,a_{D}(L_{i})|L_{2},L_{3})-\frac{2L_{1}}{\sqrt{\pi}}g\left(-\frac{1}{2},a_{D}(L_{i})\Bigg|L_{1}\right)
-\frac{2L_{2}}{\sqrt{\pi}}g\left(-\frac{1}{2},a_{D}(L_{i})\Bigg|L_{2}\right)\nonumber\\
&-&\frac{2L_{3}}{\sqrt{\pi}}g\left(-\frac{1}{2},a_{D}(L_{i})\Bigg|L_{3}\right)
-\frac{2L_{1}L_{2}L_{3}}{\pi^{\frac{3}{2}}}g\left(\frac{1}{2},a_{D}(L_{i})\Bigg|L_{1},L_{2},L_{3}\right)-a_{D}^{-1}(L_{i})(L_{1}+L_{2}+L_{3})\Bigg]\;.
\end{eqnarray}
By exploiting the asymptotic expressions (\ref{74a}), (\ref{74e}) and the following one \cite{fucci15}
\begin{equation}\label{93}
g_{1}\left(-\frac{1}{2},a_{D}(L_{i})\Bigg|L_{j}\right)\sim \sum_{n=0}^{\infty}\frac{(-1)^{n}L{_j}^{n+1}}{n!}\Gamma\left(-\frac{1}{2}-n\right)(a_{D}(L_{i}))^{2n}\zeta_{R}(-2n-1)
-\frac{\sqrt{\pi}}{2}a_{D}^{-1}(L_{i})+\frac{\sqrt{\pi}}{2L_{j}}a_{D}^{-2}(L_{i})\;,
\end{equation}
one can obtain, as $a_{D}(L_{i})\to 0$, the relation
\begin{equation}\label{94}
a_{D}^{2}(L_{i})\simeq a^{2}-\frac{\lambda}{8\pi L_{1}L_{2}L_{3}}(L_{1}L_{2}+L_{1}L_{3}+L_{2}L_{3})\ln [a_{D}(L_{i})]\;.
\end{equation}

For Neumann boundary conditions we use, as before, the relation
\begin{equation}
a_{N}^{2}(L_{i})=a^{2}+\delta a_{N}^{2}(L_{i})\;.
\end{equation}
According to (\ref{64}), the expressions for $\delta a_{N}^{2}(L_{i})$ can be obtained from the ones for $\delta a_{D}^{2}(L_{i})$ by suitably changing the
sign of specific terms. By performing the same calculations that led us to the results for $\delta a_{D}^{2}(L_{i})$ and by changing the sign, where appropriate,
one can show that, for $a_{N}(L_{i})\to 0$,
\begin{equation}\label{95}
a_{N}^{2}(L_{i})\simeq a^{2}+\frac{\lambda}{4L_{1}L_{2}}a_{N}^{-1}(L_{i})\;,
\end{equation}
when $D=3$ and $d=2$,
\begin{equation}\label{96}
a_{N}^{2}(L)\simeq a^{2}-\frac{\lambda}{16\pi L}\ln \left[ a_{N}(L)\right]\;,
\end{equation}
for $D=3$ and $d=1$, while for $D=d=3$, one finds
\begin{equation}\label{97}
a_{N}^{2}(L_{i})\simeq a^{2}+\frac{\lambda}{8 L_{1}L_{2}L_{3}}a_{N}^{-2}(L_{i})\;.
\end{equation}
We would like to point out that, even in the case of Dirichlet and Neumann boundary conditions imposed on the $d$-dimensional subspace,
the equations determining the mass parameters $a_{D}(L_{i})$ and $a_{N}(L_{i})$ represent implicit equations for the $L_{i}$-dependent critical temperature $T_{c}(L_{i})$.
As it is to be expected also in the cases of Dirichlet and Neumann boundary conditions, $T_{c}(L_{i})\to T_{c}$ whenever $L_{i}\to\infty$.
For all boundary conditions considered here and for all $d=\{1,2,3\}$, we would like to remark that although the equations found for $a^{2}_{j}(L_{i})$
cannot be, in general, solved analytically they do always possess a solution as one can verify from a qualitative analysis of the equations for $a^{2}_{j}(L_{i})$.

\section{Conclusions}

In this work we have analyzed the one-loop correction to the mass of the $\lambda\phi^{4}$ theory in a $D$-dimensional
Euclidean space containing a $d$-dimensional compact subspace. In particular, we have focused on a self-interacting
scalar field obeying Dirichlet, Neumann or periodic boundary conditions on the $d$-dimensional subspace. In order
to study the one-loop effective action of the theory we have employed the spectral zeta function regularization method.
Although the spectral zeta functions corresponding to the three boundary conditions considered here differ from each other in
their functional form, they are all written in terms of appropriate and well-known Epstein zeta functions.
From the one-loop effective action we have computed the one-loop corrections to the mass of the field. Since the
scalar field propagates in a space containing a $d$-dimensional subspace on which the field is endowed with specific
boundary conditions, the one-loop corrections to the mass of the field depend explicitly on the size of the $d$-dimensional
compact subspace. The expressions found for the one-loop corrections to the mass in the ambit of the spectral zeta
function method, need to be renormalized. While the renormalization for periodic boundary conditions can be simply performed
by requiring that all the ultraviolet divergent terms vanish when the ``lengths" of the subspace are allowed to go to infinity,
we found that for the case of Dirichlet and Neumann boundary conditions a more appropriate renormalization method is based
on the heat kernel asymptotic expansion. This procedure has led us to the general expressions for the one-loop mass correction
in (\ref{34}) through (\ref{36}). After computing the coefficients of the heat kernel asymptotic expansion for Dirichlet, Neumann,
and periodic boundary conditions, we presented explicit expressions for the one-loop correction to the mass for the different
boundary conditions in (\ref{60}) through (\ref{64}). As an application of the results found in this work, we have
considered the Ginzburg-Landau model. More precisely, we have analyzed how the critical temperature, at which the phase transition occurs,
is modified by the presence of a $d$-dimensional compact subspace. We found that in a three dimensional Euclidean space
with a $d\leq 3$ dimensional subspace our method leads to implicit equations for the modified critical temperature. Although
these equations cannot be solved explicitly in general, they do always possess a solution.

To the best of our knowledge, the results for the one-loop mass correction to the $\lambda\phi^{4}$ theory endowed with Dirichlet
and Neumann boundary conditions have not been previously presented in the literature and, hence, appear to be new. Although the case of periodic boundary conditions
is very well-known, we have decided to include it here for completeness. One of the reasons the periodic case is overwhelmingly
represented in the literature is, perhaps, due to the fact that the spectral zeta function associated with the periodic boundary condition
reduces to a single multidimensional Epstein zeta function. This contributes to the simplicity, when compared to the cases of Dirichlet
and Neumann boundary conditions, of the expression for the one-loop effective action for periodic boundary conditions (see e.g. (\ref{19}), (\ref{20}), and (\ref{21})).
The analysis performed in this work is suitable for a number of generalizations. In fact, it would be very interesting to extend the results obtained here
to more general self-adjoint boundary conditions. This would allow us to explore how different boundary conditions imposed on the field in a
compact subspace influence the symmetry breaking mechanism for the $\lambda\phi^{4}$ theory.
An additional and important study, which would complement the analysis performed in this work, consists of obtaining
expressions for the one-loop mass correction in the cases of Dirichlet and Neumann boundary conditions in the massless case.
This investigation would shed some light on how different boundary conditions influence the phenomenon of
topological mass generation. This appears to be quite an interesting question and we hope to report on this subject in a future work.\\[.2cm]

\noindent
{\bf Acknowledgment:} KK was supported by the Baylor University Summer Sabbatical Programme.

\end{document}